\documentclass[10pt,letterpaper,twocolumn]{article}	

\usepackage{graphicx}

\usepackage{epstopdf}

\begin{document}

\title{Large Scale $B-$Field in Stationary Accretion Disks}

\author{G.S. Bisnovatyi-Kogan\\
Space Research Institute, Russian\\
Academy of Sciences, Moscow, Russia;\\
 gkogan@mx.iki.rssi.ru
\and
R.V.E. Lovelace \\
Departments of Astronomy and  Applied and
Engineering Physics, \\
Cornell University, Ithaca, NY 14853-6801; \\
RVL1@cornell.edu}

\maketitle

\begin{abstract}

 We reconsider the problem of the formation of a large-scale 
magnetic field in the accretion disks around black holes.
  In contrast with previous work we  
take into account the nonuniform vertical
structure of the disk. 
The high electrical conductivity of  the outer layers of 
the disk prevents the outward diffusion of the magnetic field.
  This implies a stationary state with a strong magnetic field in
the inner parts of the accretion disk close to the black hole.

\end{abstract}

\noindent{keywords:  accretion, accretion disks --- magnetohydrodynamics ---
black hole physics}

\section{Introduction}

   Early work on disk accretion to a black hole
argued that a large-scale magnetic field of, for example, the
interstellar medium would be dragged inward and greatly compressed
by the accreting plasma (Bisnovatyi-Kogan \& Ruzmaikin 1974, 1976;
Lovelace 1976).
 Subsequently, analytic models of the
field advection and diffusion in a turbulent disk suggested, that
the large-scale field diffuses outward rapidly 
(Lubow, Papaloizou, \& Pringle 1994; Lovelace, Romanova,
\& Newman 1994)
and prevents a significant amplification of the external poloidal field by
electrical current in the accretion disk.
    This has led to the suggestion that special conditions
(non-axisymmetry) are required for the field to be advected
inward (Spruit \& Uzdensky 2005).

   We reconsider the question of the advection/diffusion
of a large-scale magnetic field in a turbulent plasma accretion
disk, taking into account its nonuniform vertical
structure. 
   The high electrical conductivity of the surface layers
of the disk, where the turbulence is suppressed
 by the radiation flux and the relatively high magnetic field,  prevents
outward diffusion of the magnetic field.
This leads in general to  a strong magnetic field in
 the inner parts of accretion disks around black holes.

\section{The fully turbulent model}

   There are two limiting accretion disk models which have analytic
solutions for a large-scale magnetic field structure. 
   The first was constructed by Bisnovatyi-Kogan and Ruzmaikin (1976) for
a stationary non-rotating accretion disk. 
   A stationary state in
this disk (with a constant mass flux onto a black hole) is
maintained by the balance between magnetic and gravitational
forces, and thermal
balance (local) is maintained by Ohmic heating and radiative
radiative conductivity for an optically thick conditions.
   The mass flux to the black hole in the accretion disk is determined by
the finite conductivity of the disk matter and the diffusion of
matter across the large-scale magnetic field as sketched
in Figure 1. 
 The value of the large-scale magnetic 
field in stationary conditions is
determined by the accretion disk mass, which in turn is
determined by the magnetic diffusivity of the matter. 
 For a laminar disk with  Coulomb conductivity (which is
very large), the mass of
the stationary disk is also very
large making the disk's self-gravity important.
   Correspondingly, the
magnetic field needed to support a mechanical equilibrium
is also very large, reaching in the central parts of
the disk $\sim
10^{11}$ G for a stellar mass black hole, for a temperature and
density at infinity
$T_\infty \sim 10^4$ K and for $\rho_\infty \sim 10^{-24}$ g/cm$^3$.
The stationary magnetic field increases with the black hole mass as $\sim
M_{bh}^{3/2}$, and with a mass flux $\sim \dot M^{3/2}$, where $\dot M \sim
\rho_\infty T_\infty^{-3/2}$  (Bisnovatyi-Kogan and
Ruzmaikin 1976).

\begin{figure}[hbtp]
\centerline{\includegraphics[scale=0.7]{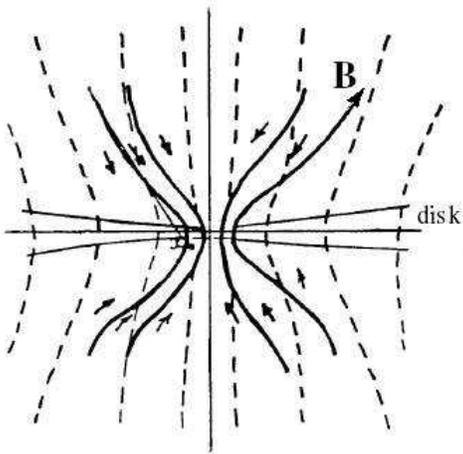}}
\caption{Sketch of the poloidal magnetic field threading an accretion
disk (from Bisnovatyi-Kogan \& Ruzmaikin 1976).
  The field strength increases with decreasing radius
 owing to flux freezing in the accreting
disk matter.} 
\label{fig5}
\end{figure}

   It is widely accepted that the laminar disk is unstable to
different hydrodynamic, magnetohydrodynamic, and plasma
instabilities which implies that the disk is turbulent. 
  In X-ray binary systems the assumption about turbulent accretion
disk is necessary for construction of a realistic models
(Shakura and Sunyaev 1973). 
    Therefore the turbulent accretion
disks had been constructed also for non-rotating models with a
large-scale magnetic field. 
   A turbulent magnetic diffusivity was
considered by Parker (1971), and by Bisnovatyi-Kogan and Ruzmaikin (1976). 
   In the last paper the turbulent diffusivity 
was scaled by the parameters of the
turbulent motion, similar to the scaling of the shear $\alpha$-viscosity
in turbulent accretion disk in binaries (Shakura \&
Sunyaev, 1973) where the viscous stress tensor component
$t_{r\phi}=\alpha P$, with $\alpha \leq 1$ a dimensionless
constant and $P$ the pressure in the disk midplane. 
In a more consistent representation, 
the coefficient of turbulent kinematic
viscosity $\nu$ in the Navier-Stokes equation is taken in the
form $\nu=(2/3) \alpha v_s h$, where $v_s=\sqrt{P/\rho}$ is the isothermal
sound speed and $\rho$ is the midplane density of the disk.
  Using this
representation, the expression for the turbulent electrical
conductivity $\sigma_t$ is
\begin{equation}
\label{eq1} \sigma_t=\frac{c^2}{\tilde\alpha 4 \pi h \sqrt{P/\rho}}.
\end{equation}
Here, $\tilde\alpha = \alpha_1 \alpha_2$. 
  The characteristic
turbulence scale is
$\ell=\alpha_1 h$, where $h$ is the
half-thickness of the disk, the characteristic turbulent velocity
is $v_t = \alpha_2 \sqrt{P/\rho}$.  
The mass of the turbulent magnetized disk is 
orders of magnitude less that in the
laminar disk case.

   The evolution of a large-scale magnetic field threading
a turbulent Keplerian disk can be estimated easily.
   This field arises from two 
sources: external electrical currents and currents in the accretion
disk.
   Evidently, the field generated by the currents in
the disk can be much
larger than that due to the external currents. 
   The magnetic field may become
dynamically important, influencing the accretion disk structure
and leading to powerful jet formation, only if it is strongly amplified
during the radial inflow of the disk matter.
This amplification is possible only when the radial accretion speed of
matter in the disk is larger than the outward diffusion speed of the
poloidal magnetic field due to the turbulent diffusivity 
$\eta_t=c^2/(4\pi \sigma_t)$.
  Estimates by Lubow et al. (1994)  shown
that for a turbulent conductivity (\ref{eq1}), 
the outward diffusion
speed is larger than the accretion speed. 
  Thus it appears that there is no large-scale
magnetic field amplification during Keplerian disk accretion. 
This conclusion is discouraging because the most plausible models of many
phenomena observed in the the systems with black holes inside the
galaxy, as well as in the extragalactic supermassive black holes
are connected with a large values of a large-scale magnetic fields.
Nevertheless, this result directly follows from the equations of the
standard disk structure, with the turbulent electric conductivity
(\ref{eq1})

Lubow et al. (1994) did numerical calculations for a simplified
situation with constant relative disk thickness $h/r$, 
constant kinematic viscosity $\nu$, and
turbulent conductivity $\sigma_t$. 
   It is easy to show that the same
result follows analytically for the standard accretion disk
structure which can be written as
$$
 \dot M = 4\pi \rho v_r rh~,\quad h={v_s}/{\Omega_K}~,\quad
 v_s=\sqrt{{P}/{\rho}}~, 
$$
\begin{equation}
\label{eq2}
4\pi r^2 h\alpha P=\dot M(j-j_{in})~,\quad
\frac{3}{2}{\Omega_K}\alpha P
 h=\frac{2aT^4c}{3\kappa \rho h}~,
\end{equation}
where $v_K=r\Omega_K$ is the Keperian velocity 
(e.g., Bisnovatyi-Kogan \& Lovelace
2001).
For regions far from the inner disk boundary,
the specific angular momentum $j=rv_K \gg j_{in}$. 
The characteristic time $t_{visc}$ of the accretion disk matter
advection  due to the shear viscosity is  $t_{visc}=r/v_r$.
From the first three relations in equation (\ref{eq2}), we obtain
\begin{equation}
\label{eq3} t_{visc}=\frac{r}{v_r}=\frac{j}{\alpha v_s^2}~.
\end{equation}
To estimate the time-scale  of outward magnetic field diffusion, we use
\begin{equation}
\label{eq4} t_{diff}=\frac{r^2}{\eta}\frac{h}{r} \frac{B_z}{B_r}~,
\end{equation}
(Lubow et al. 1994), where $B_r$ and $B_z$ are the large scale
field components evaluated at the top surface of the disk.
Here the coefficient of the magnetic turbulent diffusivity $\eta$ is
obtained from (\ref{eq1})
\begin{equation}
\label{eq5} \eta=\frac{c^2}{4\pi\sigma_t}=\tilde\alpha h v_s.
\end{equation}
  For stationary
conditions, the large-scale magnetic field in the accretion
disk is determined by the equality $t_{visc}=t_{diff}$ which implies
\begin{equation}
\label{eq6} 
\frac{B_r}{B_z}=\frac{\alpha}{\tilde\alpha}
\frac{v_s}{v_K}= \frac{\alpha}{\tilde\alpha} \frac{h}{r}\ll 1~.
\end{equation}
  In contrast, the coronal poloidal field solutions typically have
$B_r/B_z \sim 1$ at the disk surface  (Bisnovatyi-Kogan and Blinnikov 1972; 
Ustyugova et al. 1999), which implies that $t_{diff} \ll t_{visc}$.
 This inequality
indicates that the magnetic field is not amplified during
accretion for these physical conditions.

\section{Turbulent disk with radiative outer zones}

  Near the surface of the disk, in the 
region of low optical depth, the turbulent
motion is suppressed by the radiative flux, similar to the suppression of the
convection over the photospheres of stars with outer convective zones.
  The presence of the outer radiative layer does not affect the
estimate of the characteristic time $t_{visc}$ of the matter
advection in the accretion disk because it is determined by the
main turbulent part of the disk. 
  The time of the field diffusion, on the 
contrary, is significantly changed, because the electrical current is
concentrated in the radiative highly conductive regions, which
generate the main part of the magnetic field. 
  The structure of the
magnetic field with outer radiative layers is shown schematically in
Figure 2.

\begin{figure}[hbtp]
\centerline{\includegraphics[scale=0.45]{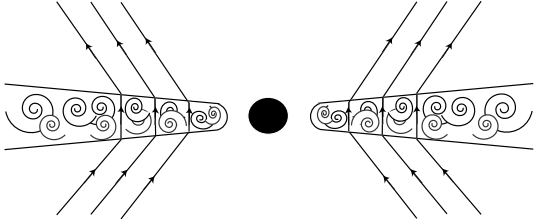}}
\caption{Sketch of the large-scale poloidal magnetic field threading a
rotating turbulent accretion disk with a radiative outer boundary
layer. 
  The toroidal current flows mainly in the highly
conductive radiative layers. 
  The large-scale (average)
field in the turbulent region is almost vertical.} 
\label{fig6}
\end{figure}

 Inside the turbulent disk the electrical current is
negligibly small so that the magnetic field there is almost fully
vertical, with $B_r \ll B_z$, according to (\ref{eq6}). 
    In the outer radiative layer, the field 
diffusion is very small, so that matter
advection is leading to strong magnetic field amplification. The
field amplification will last until the magnetic forces in the
region over the photosphere become of the order of the gravitational
ones, and start to participate in the equilibrium balance. In such
conditions the MHD and plasma instabilities are developed,
decreasing the effective electrical conductivity. We suppose, that
in the stationary state the magnetic forces could support the
optically thin regions against gravity. In the nonrotating
magnetized disk magnetic forces support the whole disk against the
gravity, so they should be much higher.
  When the magnetic force balances the gravitational force on the
outer optically thin part of the disk of
surface density  $\Sigma_{ph}$ one finds  
the following relation takes place
\begin{equation}
\label{eq8} \frac{GM\Sigma_{ph}}{r^2}\simeq \frac{B_z
I_\phi}{2c}\simeq \frac{B_z^2}{4\pi}~,
\end{equation}
(Bisnovatyi-Kogan \& Ruzmaikin
1976).
The surface density over the photosphere corresponds to a layer
with effective optical depth close to $2/3$ (e.g., Bisnovatyi-Kogan
2001). 
  We estimate the lower limit of the magnetic field strength,
taking $\kappa_{es}$ (instead of the effective opacity
$\kappa_{eff}=\sqrt{\kappa_{es}\kappa_a}$). Writing
\begin{equation}
\label{eq9} \kappa_{es}\Sigma_{ph}=2/3~,
\end{equation}
we obtain $\Sigma_{ph}=5/3$ (g/cm$^2)$ for the opacity of the Thomson
scattering, $\kappa_{es}=0.4$ cm$^2$/g. 
  The absorption opacity $\kappa_a$ is
much less than $\kappa_{es}$ in the inner regions of a luminous
accretion disk.  Thus using in equation (\ref{eq8}) $\Sigma_{ph}$ from
equation (\ref{eq9}), we estimate the lower bound on the large-scale
magnetic field of a Keplerian accretion disk as
\begin{equation}
\label{eq10}
B_z=\sqrt{\frac{5\pi}{3}}\frac{c^2}{\sqrt{GM_\odot}}\frac{1}{x\sqrt{m}}\simeq
10^8{\rm G} \frac{1}{x\sqrt{m}}~,
\end{equation}
where $x={r}/{r_g}$ and
$m={M}/{M_\odot}$.
  For comparison, the surface density $\Sigma_d$ of the
disk in the inner radiation dominated region, where we may
expect the largest values of the magnetic fields, is  
$$
\Sigma_d=\frac{80\sqrt{2}}{9\alpha}\frac{x^{3/2}}{\dot
m}\left(1-\sqrt{\frac{3}{x}}\right)^{-1}~,
$$
\begin{equation}
\label{eq11} 
\dot m=\frac{\dot M c^2}{L_c}~,\quad \quad L_c=\frac{4\pi
cGM}{\kappa_{es}}~,
\end{equation}
(see Bisnovatyi-Kogan 2001).
The maximum magnetic field is reached when the outward magnetic force
balances the gravitational force on the disk of surface mass
density $\Sigma_{ph}$.
In equilibrium, $B_z \sim \sqrt{ \Sigma_{ph}}$. 
   We find that $B_z$ in a
Keplerian accretion disk is about $20$ times less than its maximum
possible value for $x=10,\,\,\alpha=0.1,$ and  $\dot m=10$.

\section{Discussion}

An important question is the energy 
density of the large-scale stationary magnetic field in comparison with 
the rotational or gravitational energy density of the disk.
  For a non-rotating magnetized
accretion disk, the energy density of the  field is of the
order of the gravitational one so that the magnetic field strength is
very large in the vicinity of a black hole (Bisnovatyi-Kogan \&
Ruzmaikin 1976), and may be many order of magnitude larger than the
external seed field. In the case of the fully turbulent Keplerian
disk the poloidal magnetic field tends to drift outward (Lubow et
al. 1994; Lovelace et al. 1994) so that its value 
cannot significantly exceed the strength of
the  large-scale seed magnetic field.

 The suggestion of a fully turbulent accretion disk with a
small turbulent conductivity is violated in the outer surface layer
of the accretion disk where the optical depth is small and the
turbulence is suppressed by the strong radiative flux. 
   This is similar to the radiative layer above the main
body of a convective star  (similar to the
Sun; e.g., Bisnovatyi-Kogan 2001).
   In the radiative layer,
the magnetic field diffusion is much slower than in the region of fully
developed turbulence.
   In the radiative layer the diffusion is determined by the
classical Coulomb conductivity which is very large. The
diffusion in this layer is  practically negligible. 
  The electric current is concentrated in the radiative layer, and the main
body of the turbulent disk is almost current-free and thus force-free. 
The magnetic field lines in this region are almost straight
as shown in Figure 2.
  Because of the negligible diffusion in
the radiative layer, the large-scale field drifts inward
until the dynamical action of the magnetic
field on the photosphere becomes comparable with that of
centrifugal and gravitational forces. 
   At this point the inward drift of the field will
be halted and
a the stationary state formed where the
the magnetic, centrifugal, and gravitational forces on the
optically thin region will be comparable,  and
deviations from the Keplerian angular velocity  will be 
significant.
  The strength of the magnetic field for such conditions
is smaller than in the  nonrotating disk of Bisnovatyi-Kogan and
Ruzmaikin (1976), but still it is very large in the
vicinity of the black hole. 
  In this situation we may expect a
nonuniform distribution of the angular velocity over the disk
thickness:  The main body of the turbulent disk is rotates with the
velocity close to the Keplerian one, and outer optically thin layers
rotate substantially slower.

   Self-consistent models of the rotating
accretion disks with a large-scale magnetic field requires solution
the equations of magnetohydrodynamics.
   We expect two different self-consistent
solutions for the same set of the input parameters. 
   In the case of
a fully turbulent disk without radiative surface layers,
the large-scale magnetic field will remain close to
the value of the seed field and the disk's angular
velocity will be close to Keplerian. 
 In the second
solution the strength of the magnetic field is large,
and it  may greatly  exceed the strength of the seed field. 
  In this solution
the angular velocity distribution may deviate
considerably from the Keplerian law. 
    In presence of the radiative
layers the solution with a small field will not be stationary,
 and a transition to the strong field solution will take place. 
We conclude therefore that the strong field solution is the only
stable stationary solution for a rotating accretion disk.
   Further investigation of the build up of strong large-scale
fields by accretion is underway (Rothstein et al. 2007).

\section*{Acknowledgements}

 The work of G.B.-K. was partially
supported by RBFR grants 05-02-17697, 06-02-90864, and RAN program P-04.
The work of R.L. was supported in part by NASA grants NAG5-13220,
NAG5-13060, by NSF grant AST-0507760.

\end{document}